\documentclass[12pt]{iopart}
\usepackage{amssymb}
\usepackage{iopams}  
\usepackage{graphicx}
\usepackage{color} 
\usepackage{epsfig}
\definecolor{darkgreen}{rgb}{0,0.5,0} 
\definecolor{violet}{rgb}{0.5,0,0.5}
\definecolor{orange}{rgb}{0.2,0.5,0.5}
 
\newcommand{\text}[1]{{\sf #1}}
 
\newcommand{\Tt}{\mathcal{T}} 
\newcommand{\Rr}{\mathcal{R}} 
\newcommand{\Pp}{\mathcal{P}} 
\newcommand{\Ss}{\mathcal{S}} 

\begin{document}

\title[Neutral evolution in social dilemmas]{The edge of neutral evolution in social dilemmas}

\author{Jonas Cremer\dag , Tobias Reichenbach\ddag , and Erwin Frey\dag  
}

\address{\dag\  Arnold Sommerfeld Center for Theoretical Physics and
  Center for NanoScience, Department of Physics,
  Ludwig-Maximilians-Universit\"at M\"unchen, Theresienstrasse~37,
  D-80333 M\"unchen, Germany}

\address{\ddag\ Howard Hughes Medical Institute and Laboratory of Sensory Neuroscience,
The Rockefeller University,
1230 York Avenue,
New York, NY 10065
U.S.A.}

\begin{abstract}The functioning of animal as well as human societies fundamentally relies on cooperation. Yet, defection is often favorable for the selfish individual, and social dilemmas arise. Selection by individuals' fitness, usually the basic driving force of evolution,  then quickly eliminates cooperators. However, evolution is also governed by fluctuations that can be of greater importance than fitness differences,  and can render evolution effectively neutral. 
Here, we investigate the effects of selection versus fluctuations in social dilemmas. 

By studying the mean extinction times  of cooperators and defectors, a variable sensitive to fluctuations, we are able to  identify and quantify an emerging `edge of neutral evolution' which  delineates regimes of neutral and Darwinian evolution. Our results reveal that cooperation is significantly maintained in the neutral regimes. In contrast, the classical predictions of evolutionary game theory, where defectors beat cooperators, are recovered in the Darwinian regimes. Our studies demonstrate that fluctuations can provide a surprisingly simple way to partly resolve social dilemmas. 
Our methods are generally applicable to estimate the role of random drift in evolutionary dynamics.
\end{abstract}



\maketitle

\section{Introduction}

Individuals of ecological communities permanently face the choice of either cooperating with each other, or of cheating~\cite{Axelrod,NowakEGT,Ratnieks,Diggle,Fehr}. While cooperation is beneficial for the whole population and essential for its functioning, it often requires an investment by each agent. Cheating is then  tempting, yielding social dilemmas where defection is the rational choice that would yet undermine the community and could even lead to ultimate self-destruction. However, bacteria or animals do not act rationally;  instead, the fate of their populations is governed by an evolutionary process, through reproduction and death. The conditions under which cooperation can thereby evolve are subject of much contemporary, interdisciplinary research~\cite{NowakEGT,axelrod-1981-211,milinski-1987-325,rockenbach-2006-444,TraulsenGroup,Szabo2,Chuang01092009}.

Evolutionary processes possess two main competing aspects. The first one is selection by individuals' different fitness, which underlies adaptation~\cite{Fisher,Wright,Maynard} and is, by neo-Darwinists, viewed as the primary driving force of evolutionary change. In social dilemmas, defectors exploit cooperators rewarding them a higher fitness; selection therefore leads to fast extinction of cooperation, such that the fate of the community mimics the rational one. A large body of work is currently devoted to the identification of mechanisms that can reinforce  cooperative behavior~\cite{NowakCooperation}, \emph{e.g.} kin selection~\cite{Hamilton64,Hamilton}, reciprocal altruism~\cite{Trivers,NowakSigmundTFT}, or punishment~\cite{Fehr,rockenbach-2006-444,hauert-2007-316}. However, the evolution of cooperation in Darwinian settings still poses major challenges. The  second important aspect of evolution are random fluctuations that occur from the unavoidable stochasticity of birth and death events and the finiteness of populations. Neutral theories emphasize their influence which can, ignoring selection, explain many empirical signatures of ecological systems such as species-abundance relations as well as species-area relationships~\cite{Volkov,Bell,Hubbell,Muneepeerakul,Muneepeerakul2}.  The importance of neutral evolution for the maintenance of cooperation has so far found surprisingly little attention~\cite{nowak-2004-428,Traulsen,Antal,TaylorEx}. 

In this Article, we introduce a general concept capable to investigate the effects of selection versus fluctuations by analyzing extinction events. We focus on social dilemmas, i.e., we study the effects of Darwinian  versus neutral evolution on cooperation \footnote{Within this paper, we use the term 'Darwinian' to signify evolutionary dynamics mainly driven by selection, as assumed within the modern synthesis of evolution.}. For this purpose, we consider a population that initially displays coexistence of cooperators and defectors, i.e., cooperating and non-cooperating individuals. After some transient time, one of both `species' will  disappear~\cite{Hubbell}, simply due to directed and stochastic effects in evolution  and because extinction is irreversible: an extinct species cannot reappear again. The fundamental questions regarding cooperation are therefore: Will cooperators eventually take over the whole population, and if not, for how long can a certain level of cooperation be maintained?

We show that the answers to these questions  depend on the influence of stochasticity. For large fluctuations, evolution is effectively neutral, and cooperation maintained on a long time-scale, if not ultimately prevailing. In contrast, small stochastic effects render selection important, and cooperators die out quickly if disfavored. We demonstrate the emergence of an `edge of neutral evolution' delineating both regimes.  


\section{Models and Theory}

\subsection{Social dilemmas}

  \begin{table}
  \caption{
  {\bf Different types of social dilemmas}. We consider a population of cooperators and defectors, and describe their interactions in terms of four parameters $\Tt,\Rr,\Ss$ and $\Pp$, see text. Depending on the payoff-differences $\Ss-\Pp$ and $\Tt-\Rr$, four qualitatively different scenarios arise.}
\begin{center}
\begin{indented}
\item[]\begin{tabular}{l||l|l}
  & $\Ss-\Pp<0 $ & $\Ss-\Pp>0$ \\
\hline\hline  $\Tt-\Rr<0$ & Coordination game & By-product mutualism \\
 \hline $ \Tt-\Rr>0$ & Prisoner's dilemma & Snwodrift game
\end{tabular}
\end{indented}
\end{center}
\label{tab_dilemmas}
\end{table}

Consider a population of $N$ individuals which are either cooperators $C$ or defectors $D$. We assume that individuals randomly  engage in pairwise interactions, whereby cooperators and defectors behave distinctly  different and thereby gain different fitness. The population than evolves by fitness-dependent reproduction and random death, i.e., a generalized Moran process~\cite{NowakEGT,Moran}, which we describe in detail in the next Subsection. Here we present the different possible  fitness gains of cooperators and defectors.

In the \emph{prisoner's dilemma}   a cooperator provides a benefit $b$ to another individual,  at a  cost $c$ to itself (with the cost falling short of the benefit). In contrast, a defector refuses to provide any benefit and hence does not pay any costs. For the selfish individual,  irrespective of whether the partner cooperates or defects, defection is favorable, as it avoids the cost of cooperation, exploits cooperators, and ensures not to become exploited.   However, if all individuals act rationally and defect, everybody is, with a gain of $0$, worse off compared to universal cooperation, where a net gain of $b-c$ would be achieved.  The prisoner's dilemma therefore describes, in its most basic form,  the fundamental problem of establishing cooperation.

We can generalize the above scheme to include other basic types of social dilemmas~\cite{Dawes}. Namely, 
two cooperators that meet are both rewarded a payoff $\Rr$, while two defectors obtain a punishment $\Pp$. When a defector encounters a cooperator, the first exploits the second, gaining the temptation $\Tt$, while the cooperator only gets the suckers payoff $\Ss$. Social dilemmas occur when $\Rr>\Pp$, such that cooperation is favorable in principle, while temptation to defect is large: $\Tt>\Ss,~\Tt>\Pp$. These interactions may be summarized  by the payoff matrix 
\begin{displaymath}
 \label{eq:payoff}
 \begin{array}{l|ll}
  & C  & D \\ \hline
 C& \Rr & \Ss \\
 D & \Tt  & \Pp
 \end{array}
 \end{displaymath}
Hereby, the entries in the upper row describe the payoff that a cooperator obtains when encountering a cooperator $C$ or a defector $D$, and the entries in the lower row contain the payoffs for a defector.
  
\begin{figure}
\begin{center}
\includegraphics[width=12cm]{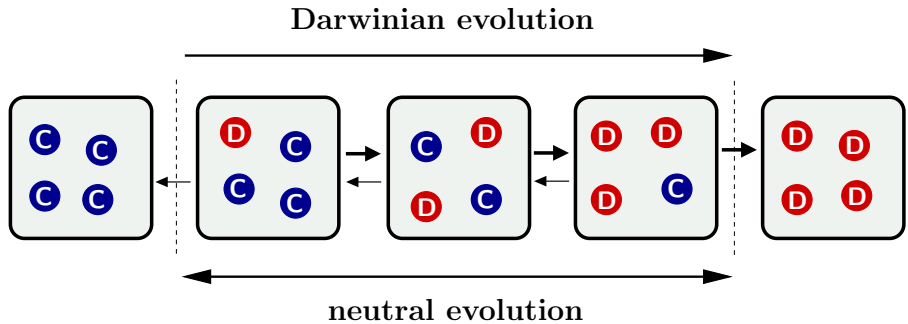}
\end{center}
\caption{{\bf Phase space exemplified for the prisoner's dilemma.} The evolutionary dynamics  consists of a Darwinian, directed part caused by selection of defectors (D) against cooperators (C), and a neutral, undirected part  due to fluctuations.  Eventually, only one species survives.\label{fig:phasespace}}
\end{figure}
 Variation of the parameters $\Tt,\Pp,\Rr$ and $\Ss$ yields  four principally different types of games, see Tab.~\ref{tab_dilemmas} and Fig.~\ref{fig:C}.  The \emph{prisoner's dilemma} as introduced above arises if the temptation $\Tt$  to defect is  larger than the reward $\Rr$, and if the punishment $\Pp$ is larger than the suckers payoff $\Ss$, e.g., $\Rr=b-c$, $\Tt=b$, $\Ss=-c$ and $\Pp=0$. As we have already seen above, in this case, defection is the best strategy for the selfish player. Within the three other types of games, defectors are not always better off. For the \emph{snowdrift game} the temptation $\Tt$ is still higher than the reward $\Rr$ but the sucker's payoff $\Ss$ is larger than the punishment $\Pp$. Therefore,  cooperation is favorable when meeting a defector, but defection pays off when encountering a cooperator, and a rational strategy consists of a mixture of cooperation and defection.   Another scenario is the \emph{coordination game}, where  mutual agreement is preferred: either all individuals cooperate or defect as the reward $R$ is higher than the temptation $\Tt$ and the punishment $\Pp$ is higher than sucker's payoff $\Ss$. Last, the scenario of \emph{by-product mutualism} yields cooperators fully dominating defectors since the reward $\Rr$ is higher than the temptation $\Tt$ and the sucker's payoff $S$ is higher than the punishment $\Pp$. All four situations and the corresponding ranking of the payoff values are depicted in Tab.~\ref{tab_dilemmas}  and Fig.~\ref{fig:C}.

\subsection{The evolutionary dynamics}

We describe the evolution by a generalized Moran process~\cite{NowakEGT,Moran}, where the population size $N$ remains constant and reproduction is fitness-dependent, followed by a random death event.

Let us  denote the number of cooperators by $N_C$; the number of defectors then reads $N-N_C$.  The individuals' fitness are given by a constant background fitness, set to $1$, plus the payoffs obtained from social interactions. The fitness of cooperators and defectors thus read $f_C=1+\Rr (N_C-1)/(N-1)+\Ss(N-N_C)/(N-1)$ and $f_D=1+\Tt N_C/(N-1)+\Pp(N-1-N_C)/(N-1)$, resp..  In the following, we assume weak selection, i.e., the payoff coefficients are small compared to the background fitness. Note that within this limit, the self interactions of individuals are only of minor relevance. More important, in the case of weak selection, the evolutionary dynamics of the game depends only on the payoff differences  $\Tt-\Rr$ and $\Ss-\Pp$. The different types of social dilemmas arising from theses two parameters are listed in Table 1.

\begin{figure}
\begin{center}
\hspace{-5.7cm}{(A)}\hspace*{4.2cm}{(B)}\hspace*{5cm}\\\vspace*{0.3cm}
\includegraphics[height=5.5cm]{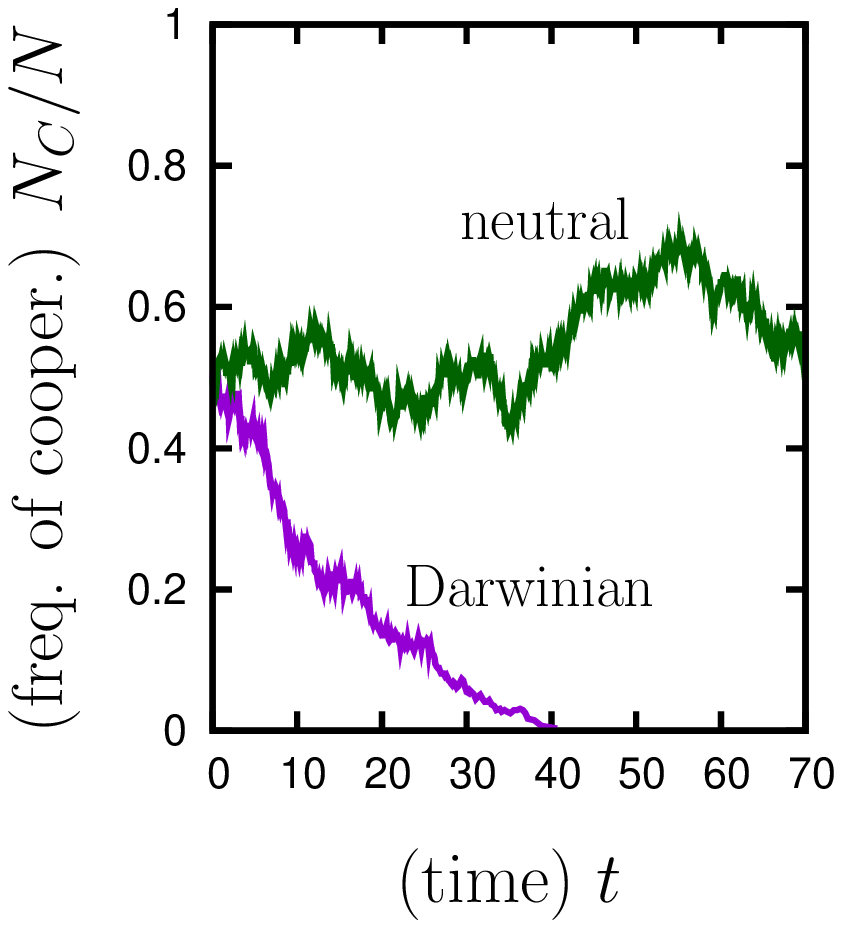}
\includegraphics[height=5.5cm]{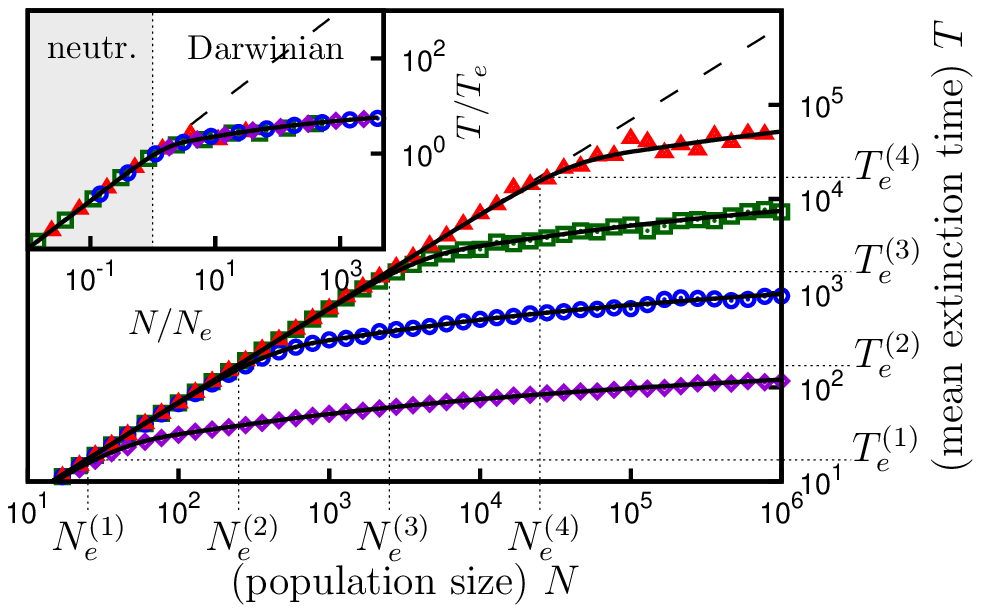}
\end{center}
\caption{{\bf The prisoner's dilemma.} Defectors save the  cost $c$ of cooperation and therefore have a fitness advantage of $c$  compared to cooperators.  (A), Exemplary evolutionary trajectories. A high selection strength, i.e., a high fitness difference $c=0.1$ (purple), leads to Darwinian evolution and fast extinction of cooperators, while a small one, $c=0.001$ (green), allows for dominant effects of fluctuations and maintenance of cooperation on long time-scales. We have used $N=1000$ in both cases.  (B), The dependence of the corresponding mean extinction time $T$ on the system size $N$. We show data from stochastic simulations as well as analytical results (solid lines) for  $T$, starting from equal abundances of both species, for different values of $c$ (see text): $c_1=0.1$ (\textcolor{violet}{$\diamondsuit$}), $c_2=0.01$ 
(\textcolor{blue}{\Large $\circ$}), $c_3=0.001$ (\textcolor{darkgreen}{\scriptsize $\Box$}), and  $c_4=0.0001$ (\textcolor{red}{\scriptsize $\bigtriangleup$}). The transition from the neutral to the Darwinian regime occurs at population sizes $N_e^{(1)}, N_e^{(2)}, N_e^{(3)}$, and $N_e^{(4)}$. They scale as $1/c$: $N_e\approx 2.5/c$, as is confirmed by the rescaled plot where the data collapse onto the universal scaling function $G$, shown in the inset.}
\label{fig:A}
\end{figure}


In the Moran process, reproduction of individuals occurs proportional to their fitness, and each reproduction event is accompanied by death of a randomly chosen individual.
As an example, the rate for reproduction of a defector and corresponding death of a cooperator reads
\begin{eqnarray}
\label{eq:transprob}
\Gamma_{C\to D}&=&\frac{f_D}{\langle f \rangle}\frac{N_C}{N}\frac{N-N_C}{N}\,,
\end{eqnarray}
whereby $\langle f\rangle=f_C N_C/N + f_D(1-N_C/N)$ denotes  the average fitness.
The time scale is such that an average number of $N$ reproduction and death events occur in one time step.

\subsection{Distinguishing Darwinian  from neutral evolution: Extinction times}

The evolutionary dynamics is intrinsically stochastic. Although defectors may have a fitness advantage compared to cooperators, the latter also have a certain probability to increase. This situation is illustrated in Fig.~\ref{fig:phasespace} for a population of $4$ individuals and the dynamics of the prisoner's dilemma. Darwinian evolution, through selection by individuals' fitness,  points to the `rational' state of only defectors, while fluctuations oppose this dynamics and can lead to a state of only cooperators. In any case, upon reaching overall defection or cooperation,  the temporal development comes to an end. One species therefore eventually dies out.   

The mean extinction time, i.e., the mean time it takes a population where different species coexist to eventually become uniform, allows to distinguish Darwinian from neutral evolution. Consider the  dependence of the mean extinction time $T$  on the system size $N$. Selection, as a result of some interactions within a finite population, can either stabilize or destabilize a species' coexistence with others as compared to neutral interactions, thereby altering the mean time until extinction occurs. 
Instability leads to steady decay of a species, and therefore to fast extinction~\cite{Antal,TaylorEx,Reichenbach}: The mean extinction time $T$ increases only logarithmically in the  population size $N$, $T\sim \ln N$, and a larger system size does not ensure much longer coexistence. This behavior can be understood by noting that a species disfavored by selection decreases  by a constant rate. Consequently, its population size decays exponentially in time, leading to a logarithmic dependence of the extinction time on the initial population size.
In contrast, stable existence of a species induces $T\sim \exp N$, such that extinction takes an astronomically long time for large populations~\cite{Antal,TaylorEx,Cremer}.  In this regime, extinction only stems from large fluctuations that are able to cause sufficient deviation from the (deterministically) stable coexistence. These large deviations  are exponentially suppressed and hence the time until a rare extinction event occurs scales exponentially in the system size $N$.

An intermediate situation, i.e., when $T$ has a power-law dependence on $N$, $T\sim N^\gamma$, signals dominant influences of stochastic effects and corresponds  to  neutral evolution~\cite{reichenbach-2006-74,berr:048102,traulsen_extime}. Here the extinction time grows considerably, though not exponentially, in increasing population size.  Large $N$ therefore clearly prolongs coexistence of species but can still allow for extinction within biologically reasonable  time-scales. A typical neutral regime is characterized by $\gamma=1$,  such that $T$ scales linearly in the system size $N$.  This corresponds to the case where the dynamics yields an essentially unbiased random walk in state space. The mean-square displacement  grows linearly in time, with a diffusion constant proportional to $N$. The absorbing boundary is thus reached after a time proportional to the system size $N$. Other values of $\gamma$ can occur as well. For example, and as shown later, $\gamma=1/2$ can occur in social dilemmas (regimes (2) in Fig. \ref{fig:C}).

To summarize,  the mean extinction time $T$  can be used to classify evolutionary dynamics into a few fundamental regimes. Darwinian evolution can yield  stable and unstable coexistence, characterized by $T\sim \log N$ and $T\sim \exp N$, resp.. Power law dependences, $T\sim N^\gamma$, indicate neutral evolution.  Transitions between these regimes can occur and manifest as crossovers in the functional relation $T(N)$.

\subsection{Analytical description}

An approximate analytical description, valid for a large number $N$ of interacting individuals, is possible. The quantity of interest is thereby the  probability $P(N_C,t)$ of having $N_C$ cooperators at time $t$. Its time evolution is described by a master equation specified by transition rates such as~(\ref{eq:transprob}).   For large population sizes $N$ the master equation   can be approximately described within a generalized diffusion approach, where the fraction $x=N_C/N$ of cooperators is considered as a continuous variable. The temporal development of $P(x,t)$ is then described by a Fokker-Planck equation~\cite{ewens,Kimura,Traulsen},
\begin{eqnarray}
\label{eq:FPE}
\frac{\partial\;}{\partial t} P(x,t)=-\frac{\partial\;}{\partial x}\left[\alpha(x)P(x,t)\right]+\frac{1}{2}\frac{\partial^2\;}{\partial x^2}\left[\beta(x)P(x,t)\right]\,.
\end{eqnarray}
Hereby, $\alpha(x)$ describes the Darwinian of the evolution, due to selection by fitness differences, and corresponds to the deterministic dynamics $\frac{d}{dt} x=\alpha(x)$. The second part, which involves the diffusion term $\beta(x)$,  accounts for fluctuations (to leading order) and thereby describes undirected random drift.  $\beta(x)$ decreases like $1/N$ with increasing population size. For the social dilemmas which we study in this article $\alpha$ and $\beta$ are given by,
\begin{eqnarray}
  \label{eq:FPEcoeff}
  \alpha(x)&=&x(1-x)\left[(\Ss-\Pp)(1-x)-(\Tt-\Rr) x \right]\nonumber ,\\
  \beta(x)&=&\frac{1}{N}x(1-x)\left[2+(\Ss-\Pp)(1-x)+(\Tt-\Rr) x\right]\nonumber\\
  &\approx& \frac{2}{N}x(1-x)\,.
\end{eqnarray}
Here, the approximation of $\beta$ given in the last line is valid since weak selection is assumed.

The prisoner's dilemma, specified by  $\Tt-\Rr=\Pp-\Ss\equiv c>0$ describes the situation where defectors have a frequency independent fitness advantage $f_D-f_C=c$ as compared to cooperators. This  scenario is frequently studied in population genetics~\cite{ewens}; we briefly discuss it in the following. The directed part and diffusion coefficients are given by, 
\begin{eqnarray}
\alpha(x)&=&-c x(1-x)\,,\nonumber\\
\beta(x)&=&\frac{1}{N}x(1-x)\left[2-c(1-2x)\right]\approx\frac{2}{N}x(1-x)\,.
\end{eqnarray} 
With these one can calculate the fixation probability $P_{ \text{fix,C} }$ to end up with only cooperators if starting with an equal fraction of cooperators and defectors. It has already been calculated in previous work~\cite{Kimura,ewens} and reads,
\begin{eqnarray}
\label{eq:fixprob}
P_\text{fix,C}=\frac{e^{-Nc/2}-e^{-Nc}}{1-e^{-Nc}}\,.
\end{eqnarray}
The probability for fixation of defectors follows as $P_\text{fix,D}=1-P_\text{fix,C}$. Within the Darwinian regime ($Nc\to\infty$) defectors fixate ($P_\text{fix,D}=1)$, whereas for the neutral regime ($Nc\to 0$) both strategies have the same chance of prevailing ($P_\text{fix,C}=P_\text{fix,D}=1/2$).  

The fixation probability gives no information about the typical time needed for extinction of one of the two species. However,  this time is important to determine whether extinction happens within the time scale of observation. We turn to this question in the following.
 
\subsection{Analytical calculation of mean extinction times}

The above analytical description, in form of the Fokker-Planck equation~(\ref{eq:FPE}),  can be employed for computing the mean extinction time $T(x)$. The latter refers to the mean time required for a population initially consisting of a fraction $x=N_C/N$ of cooperators to reach a uniform state (only either cooperators or defectors). It is given as solution to the corresponding backward Kolmogorov equation~\cite{Gardiner,Risken},
\begin{equation}
\label{eq:BackwardMFPTh}
\left[ \alpha(x)\frac{\partial\;}{\partial x}+\frac{1}{2}\beta(x)\frac{\partial^2\;}{\partial x^2} \right]T(x)=-1\,,
\end{equation}
with appropriate boundary conditions. This equation can be solved by iterative integration~\cite{Gardiner}. In detail, the mean extinction time, $T=T(x=1/2)$, if starting with an equal fraction of cooperators $x=1/2$ is given by 
\begin{eqnarray}
\label{eq:integratingdetails}
T=2 &&\left[\left(\int_0^{1/2} du/\Psi(u) \right) \int_{1/2}^1 dy/\Psi(y)\int_0^y dz \Psi(z)/\beta(z)\right.\nonumber\\
&&\,\;-[\left.\left(\int_{1/2}^{1} du/\Psi(u) \right) \int_{0}^{1/2} dy/\Psi(y)\int_0^y dz \Psi(z)/\beta(z)\right]\nonumber\\ 
&&\times\left[\int_0^1 du /\Psi(u) \right]^{-1},
\end{eqnarray} 
where $\Psi(x)$ is given by $\Psi(x)=\exp\left(\int_0^x dy\, 2\alpha(y)/\beta(y) \right)$. We have performed these integrals for the general Moran process and show the results in the following. 

For the special case of the prisoner's dilemma, specified by $\Tt-\Rr=\Pp-\Ss\equiv c >0$,  (frequency independent fitness advantage), Eq.~(\ref{eq:integratingdetails}) can be solved exactly. The solution reads,
\begin{eqnarray}
T&=&\frac{1}{N}\Big[ \text{P}_\text{fix,C}\left\{-\ln(cN)-\gamma+\text{Ei}\left(cN/2\right) +e^{cN}\left[\text{Ei}\left(-cN\right) -\text{Ei}\left(-cN/2\right)\right]\right\} \Big.\nonumber\\
&&+\Big.\text{P}_\text{fix,D}\left\{\ln{(cN)}+\gamma-\text{Ei}\left(-cN/2\right)+e^{-cN}\left[\text{Ei}\left(cN/2\right)-\text{Ei}\left(cN \right) \right]\right\}\Big]\,,
\label{eq:T_constant_fitness}
\end{eqnarray}

where $\text{Ei}(x)$ denotes the exponential integral $\text{Ei}(x)=\int x^{-1}\exp(x)dx$ and $\gamma\approx 0.577$ is the Euler Mascheroni constant. $P_{\text{fix},C}$ and $P_{\text{fix},D}$ denote the fixation probabilities of cooperators and defectors, given by Eq.~(\ref{eq:fixprob}).  The analytical solution of the mean extinction time as a function of $N$ is shown and compared to stochastic simulations in Fig.~\ref{fig:A}. For a further discussion of $T(N)$ (Eq.~\ref{eq:T_constant_fitness}) and its impact on  evolutionary dynamics we defer the reader to section \ref{sec:results}. Here, just note that the asymptotic behavior, of $\text{Ei}\left(x\right)$ is given by $\text{Ei}(x)\approx \log(|x|)+\gamma+x$ for $x\to 0$, and $\text{Ei}(x)\approx \pm \log(|x|)+\exp(x)/x$ for $x\to \pm \infty$. With this, the well known asymptotic solutions for high and low population size $N$, $T=\log(2) N$ and $T\sim \log N$ are obtained. 

For general social dilemmas with arbitrary payoff values $\Tt,\Pp,\Rr,\Ss$, we need to rely on some approximations.  Using the drift and diffusion coefficient given by Eq.~(\ref{eq:FPEcoeff}) we now linearize the fraction $\alpha/\beta$, i.e. we write $\alpha(x)/\beta(x)\approx g(x-x^*)$.  Hereby $x^*=(\Ss-\Pp)/(\Ss-\Pp+\Tt-\Rr)$ denotes the fixed point of the deterministic dynamics, where $\alpha(x^*)=0$, and $g=-N(\Ss-\Pp+\Tt-\Rr)/2$. As an example, in the situation $\Ss-\Pp+\Tt-\Rr>0$, $|x^*|\gg 1/\sqrt{g}$, and $|1-x^*|\gg 1/\sqrt{g}$, we obtain the mean extinction time, 
\begin{eqnarray}
&&T=~\frac{N\text{ln}(2)}{g}
\lbrace\text{erfi}\left[\sqrt{g}\left(1-x^*\right)\right]+\text{erfi}\left(\sqrt{g}x^*\right)\rbrace^{-1}\times\nonumber\\
\times&&\Big\{\text{erfi}\left(-\sqrt{g}x^*\right)\left[-\mathcal{F}\left(g\left(1-x^*\right)\right)+\mathcal{F}\left(g\left(1/2-x^*\right)\right) \right]\Big.\nonumber \\
&& - \text{erfi}\left(\sqrt{g}\left(1-x^*\right)\right)\left[\mathcal{F}\left(g\left(1/2-x^*\right)\right)-\mathcal{F}\left(g\left(x^*\right)\right) \right]\nonumber \\
&& + \Big. \text{erfi}\left(\sqrt{g}\left(1/2-x^*\right)\right)\left[\mathcal{F}\left(g\left(1-x^*\right)\right)-\mathcal{F}\left(g\left(x^*\right)\right)\right]\Big\}
\label{eq:T_gen}
\end{eqnarray}
Hereby, $\text{erfi}(x)=\frac{2}{\sqrt{\pi}}\int_0^x dy \exp(y^2)$ denotes the complex error function, and  $\mathcal{F}(x)\equiv x \text{F}_{1,1;3/2,2}(x)$  involves  a generalized hypergeometric function. For graphical representation of Eq.(8) see Fig.~\ref{fig:D}A (upper branch). As before, the correct asymptotic behavior can also be calculated for this case. Note that the asymptotic behavior of $\mathcal{F}(x)$ is given by $\mathcal{F}(x)\approx x$ for $x\to 0$ and $\mathcal{F}(x)\approx \text{erfi}(\sqrt{x})-\log(|x|)/2-1$ for $x\to \pm \infty$. For small population size, the mean extinction time scales again like $T=\log(2) N$. For asymptotically large system sizes the scaling depends on the value of the fixed point $x^*$. For an internal fixed point $x^*\in (0,1)$, as arises in the snowdrift game, $T$ scales as expected like $T\sim \exp(N)$.

In the Results Section, we analyze the properties of the analytical form of the mean extinction time,  Eqs.~(\ref{eq:T_constant_fitness}) and (\ref{eq:T_gen}), together with numerical simulations, and demonstrate how it defines an emerging edge of neutral evolution.

\subsection{Edges of neutral evolution}
\label{methods_edges}

In the Results section, we show that the mean extinction time, Eq.~({\ref{eq:T_gen}}), exhibits different regimes of neutral and Darwinian dynamics. Here, we provide further information on how the boundaries between these regimes can be obtained analytically. 

\begin{figure}
\centerline{\includegraphics[width=10cm]{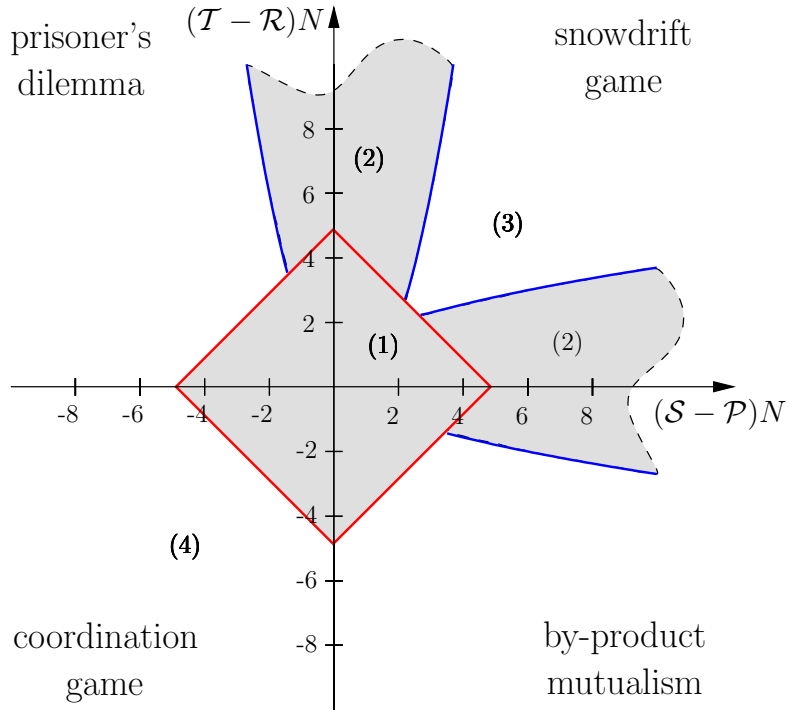}}
\caption{
{\bf Social dilemmas.} Depending on the sign of the payoff differences $\Tt-\Rr$ and $\Ss-\Pp$, a prisoner's dilemma, snowdrift game, by-product mutualism or coordination game arises. Two regimes of neutral evolution, (1) and (2) shown in grey, intervene two Darwinian regimes, (3) and (4), depicted in white. Coexistence of cooperators and defectors is lost after a mean time $T$ which discriminates the distinct regimes: In (1), we encounter $T\sim N$, while $T\sim\sqrt{N}$ emerges in (2), $T\sim\exp{N}$ in (3), and $T\sim\ln{N}$ in (4). In the prisoner's dilemma and the coordination game, neutral evolution can thus maintain cooperation at a much longer time than Darwinian evolution. The edges of neutral evolution, red and blue curves, scale as $1/N$ (see text). We therefore show them depending on $(\Tt-\Rr)N$ and $(\Ss-\Pp)N$, where they adopt universal shapes.}
\label{fig:C}
\end{figure}

For this purpose, we further approximate the dynamics. Let us, firstly, focus on the edge of the regime where $T\sim N$ emerges.  We note that, before unavoidable extinction of one species occurs, a quasi-stationary distribution may form around the fixed point $x^*$. Following the generic behavior of an Ornstein-Uhlenbeck process, its shape is approximately Gaussian \cite{Kampen}. Its width is  given by $w\sim\sqrt{1/|g|}$. $x^*$ and $g$ are specified in the preceding section. Now, for small width, $w\ll 1$, the Darwinian evolution dominates the behavior, meaning $T\sim\text{ln}(N)$ or $T\sim \exp(N)$.  In contrast, if $w\gg 1$ the dynamics is essentially a random walk, and $T\sim N$ emerges. The edge of neutral evolution therefore arises at $w\sim 1$. Remembering that $g$ is given by $g=-N(\Ss-\Pp+\Tt-\Rr)/2$, it follows that the edge between both regimes for $\Ss-\Pp,~\Tt-\Rr>$ is described by $(\Tt-\Rr)=d/N-(\Ss-\Pp)$. Numerical simulations yield a good agreement with this prediction. As discussed later (see Fig. 2), they reveal that the crossover between the two regimes is remarkably sharp. The constant $d$ which specifies the exact position of the crossover can therefore be estimated as $d\approx 5$. It follows that the regime of $T\sim N$ therefore corresponds to the square circumscribed by straight lines connecting the points $(\Tt-\Rr,\Ss-\Pp)=(5/N,0),(0,-5/N),(-5/N,0),(0,5/N)$ as shown in Fig.~\ref{fig:C}.

A similar argument allows to determine the crossover from the other neutral regime, with $T\sim \sqrt{N}$, to the Darwinian regimes. The neutral regime emerges if the fixed point $x^*$ is close to the boundaries, such that  $w\sim |x^*|$ or $w\sim |1-x^*|$ denotes the crossover to the Darwinian regimes. From these relations, if follows that the shapes of this second neutral regime are described by  $\Tt-R\approx-(\Ss-\Pp)+(\Ss-\Pp)^2 N$ and $\Ss-\Pp\approx-(\Tt-\Rr)+(\Tt-\Rr)^2 N$. The proportionality constant has again been estimated from numerical simulations. From the latter, we have also found that the parabolic curves constitute a valid approximation to this second edge of neutral evolution.

\section{Results}
\label{sec:results}
We employ the analytical expression, Eqs.~(\ref{eq:T_constant_fitness}) and (\ref{eq:T_gen}), for the mean extinction time, as well as computer simulations, to show how  regimes of Darwinian and neutral evolution can be distinguished. We demonstrate that neutral evolution can maintain cooperation on much longer time-scales than Darwinian, even if cooperation has a fitness disadvantage.

\subsection{Prisoner's dilemma}

We start with the special case of the prisoner's dilemma where defectors have a frequency independent fitness advantage $c$ compared to  cooperators. The fixation probabilities, Eq.~(\ref{eq:fixprob}), provides first insight into the dynamics.   When the population size $N$ is large and selection by fitness differences dominates the dynamics, i.e., when $cN\gg 1$,  the probability that defectors ultimately take over the whole population tends to $1$. Cooperators are guaranteed to eventually die out.  This is the regime of \emph{Darwinian} evolution; the resulting outcome equals the one of rational agents. However, in the situation of small populations and small fitness difference, i.e., $cN\ll 1$, both cooperators and defectors have an equal chance of $1/2$ of fixating. In this regime, fluctuations have an important influence and dominate the evolutionary dynamics, leaving fitness advantages without effect, evolution is \emph{neutral}. 

Further quantification of the regimes of Darwinian and neutral evolution is feasible by considering the mean extinction time, given by Eq.~(\ref{eq:T_constant_fitness}).
It is compared to stochastic simulations in Fig.~\ref{fig:A} B for different costs (fitness advantages) $c$. The excellent agreement confirms the validity of our analytic approach.  Regarding the dependence of $T$ on the population size $N$ and the fitness difference $c$, the mean extinction time can be  cast into the form,
\begin{equation}
\label{eq:one}
T(N,c)=T_e\text{G}\left(N/N_e\right),
\end{equation}
with a scaling function $\text{G}$. $T_e$ and $N_e$ are characteristic time scales and population sizes depending only on the selection strength $c$. Analyzing its properties, it turns out that $\text{G}$ increases linearly in $N$ for small argument $N/N_e\ll 1$, such that $T\sim N$, c.f.~Fig.~\ref{fig:A} B.  This is in line with our classification scheme and the expected behavior. It indicates~\cite{Antal,TaylorEx} that for small system sizes, $N\ll N_e$, evolution is \emph{neutral}.  Fluctuations dominate the evolutionary dynamics while the fitness advantage of defectors does not  give them an edge, c.f. Fig.~\ref{fig:A} A. Indeed, in this regime, cooperators and defectors have an equal chance of surviving, see Eq.~(\ref{eq:fixprob}). The $T\sim N$ behavior shows that the extinction time considerably grows with increasing population size; a larger system size proportionally extends the time cooperators and defectors coexist. 
As expected, a very different behavior emerges for large system sizes, $N/N_e\gg 1$, where $\text{G}$  increases only logarithmically in $N$, and therefore $T\sim \ln N$, again in correspondence with our classification scheme of the mean extinction time. The extinction time remains small even for large system sizes, and coexistence of cooperators and defectors is unstable.  Indeed, in this regime,  selection dominates over fluctuations in the stochastic time evolution and quickly drives the system to a state where only defectors remain, c.f.~Fig.~\ref{fig:A} A.  The evolution is \emph{Darwinian}. 

As described above, the regimes of neutral and Darwinian evolution emerge for  $N/N_e\ll 1$ and $N/N_e\gg 1$, respectively. The cross-over population size $N_e$ delineates both scenarios. Further analyzing the universal scaling function $\text{G}$, as well as comparison with data from stochastic simulations, see Fig.~\ref{fig:A} B, reveals that the transition at $N_e$ is notably sharp. We therefore refer to it as the \emph{edge of neutral evolution}.

The crossover time $T_e$ and the crossover population size $N_e$ which define the edge of neutral evolution decrease as $1/c$ in increasing  cost $c$.  This can be understood by recalling that the cost $c$ corresponds to the fitness advantage of defectors and can thus be viewed as the selection strength. The latter drives the Darwinian dynamics which therefore intensifies when $c$ grows, and the regime of neutral evolution diminishes. On the other hand, when the cost of cooperation vanishes, evolution becomes neutral also for large populations. Indeed, in this case, defectors do not have a fitness advantage compared to cooperators; both do equally well.  Our approach now yields information about how large the cost may be until evolution changes from neutral to Darwinian.  From numerical inspection of $\text{G}$ we find that neutral evolution is present for $cN < 2.5$, and Darwinian evolution takes over for $cN > 2.5$. This resembles a condition previously derived by Kimura, Ohta, and others~\cite{Kimura,KimuraFixation,Ohta} for frequency independent fitness advantages. The edge of neutral evolution arises at $N_e=2.5/c$ and $T_e=2.5/c$. 

As a consequence we note that, though selection pressure clearly disfavors cooperation, our results reveal that the ubiquitous presence of randomness (stochasticity) in any population dynamics opens a window of opportunity where cooperation is facilitated. In the regime of neutral evolution, for $cN < 2.5$, cooperators have a significant chance of taking over the whole population when initially present. Even if not, they remain on time-scales proportional to the system size, $T\sim N$, and therefore considerably longer than in the regime of Darwinian evolution, where they extinct after a short transient time, $T\sim\ln N$.

\begin{figure}
\begin{center}
\hspace*{-3cm}{(A)}\hspace{.29\textwidth} {(B)}\hspace{.29\textwidth}{(C)}\\\vspace*{0.2cm}
\hspace*{-0.5cm}\includegraphics[height=3.4cm]{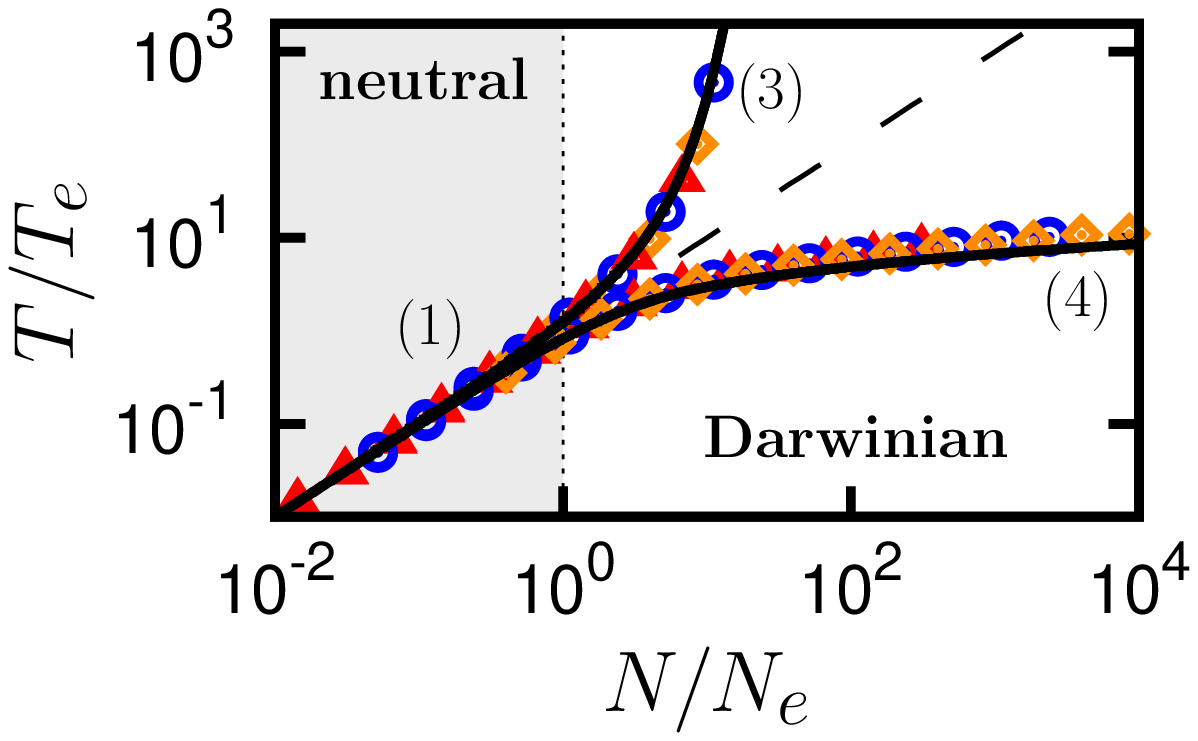}
\hspace*{.015in}
\includegraphics[height=3.4cm]{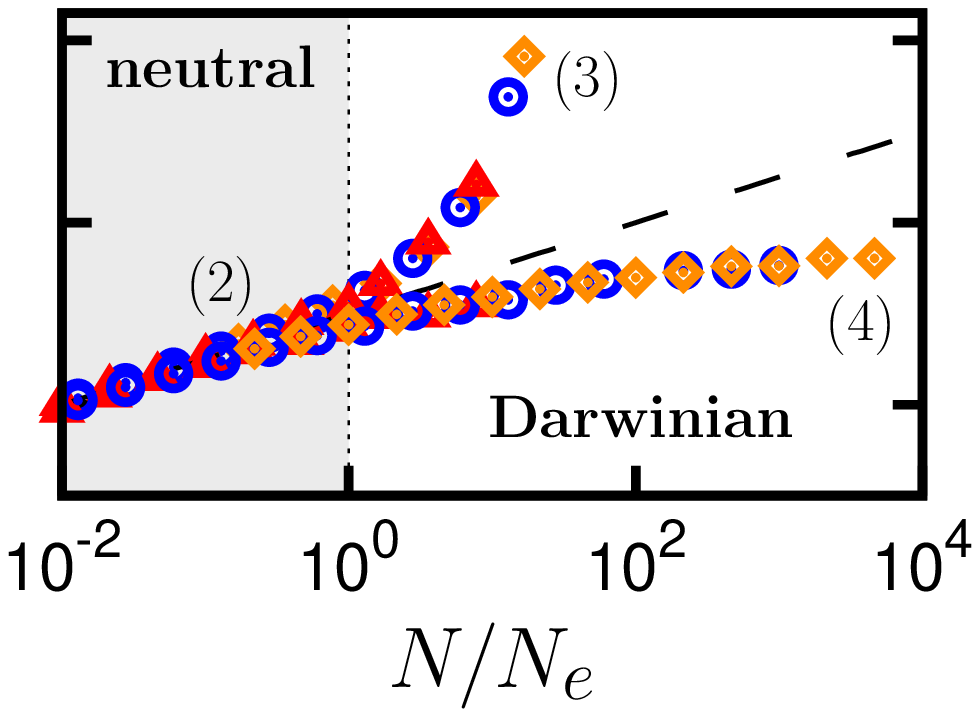}
\hspace*{.015in}
\includegraphics[height=3.4cm]{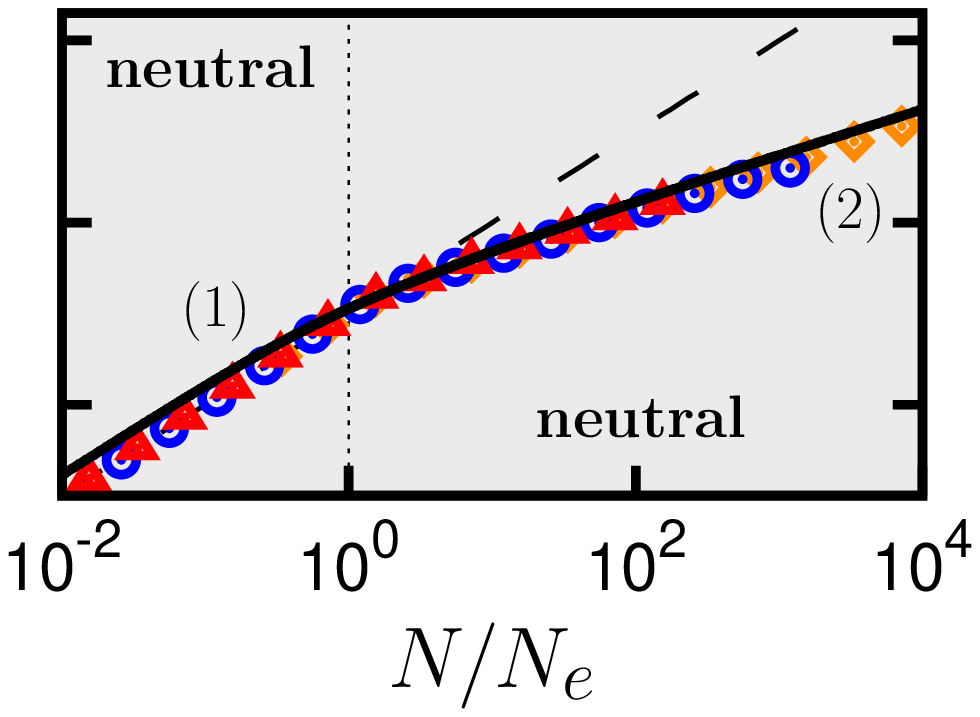}
\end{center}
\caption{{\bf Transitions and universal scaling.}  We show the rescaled mean extinction time, $T/T_e$, depending on $N/N_e$, for different transitions emerging in social dilemmas (c.f.~Fig.~\ref{fig:C}). (A), Transition from the neutral regime (1), where $T\sim N$ emerges, to the  Darwinian regimes (3) ($T\sim\exp N$) as well as (4) ($T\sim \ln N$). (B), From neutral dynamics in regime (2) ($T\sim\sqrt{N}$) to the Darwinian regimes (3) ($T\sim\exp N$) and (4) ($T\sim\ln N$). (C), Transition between the two neutral regimes (1) ($T\sim N$) and (2) ($T\sim \sqrt{N}$). Analytical calculations are shown as black lines, and symbols have been obtained from stochastic simulations for large (\textcolor{red}{\scriptsize $\bigtriangleup$}), medium (\textcolor{blue}{\Large $\circ$}), and small (\textcolor{orange}{$\diamondsuit$}) values of $\Ss-\Pp$ and/or $\Tt-R$. The data collapse onto universal curves reveals the accuracy of the scaling laws.
In (A), we have used  $\Ss-\Pp=\Tt-\Rr\in\lbrace -0.1,-0.01,-0.001,0.001,0.01,0.1 \rbrace$, while  $\Ss-\Pp\in\lbrace -0.1,-0.01,-0.001,0.001,0.01,0.1 \rbrace$, $\Tt-\Rr=1$ in (B), and $\Ss-\Pp=0$, $\Tt-\Rr\in\lbrace 0.001, 0.01, 0.1\rbrace$ in (C).}
\label{fig:D}
\vspace*{1cm}
\end{figure}

\subsection{General social dilemmas}

Let us now consider the influence of fluctuations within  the more general form of social dilemmas, given by the parameters $\Tt,\Pp,\Rr,\Ss$. We employ the analytical form of the mean extinction time,  Eq.~(\ref{eq:T_gen}), as well as results from stochastic simulations.  Examples for different paths in parameter space are shown in Fig.~\ref{fig:D}. Again, the approximative analytical results agree excellently with numerics.

Concering the dependence of the mean extinction time on the population size,  different  behaviors emerges, reflecting the different regimes of evolutionary dynamics. Two regimes of Darwinian evolution form, depicted white in Fig.~\ref{fig:C}. The first one occurs within the snowdrift game, where the extinction time increases exponentially in the population size, $T\sim\exp{N}$, and coexistence of cooperators and defectors is stable. The second regime comprises parts of the prisoner's dilemma, the coordination game, and by-product mutualism. There, either defectors or cooperators eventually survive, and the mean extinction time of the other strategy is small, and obeys a logarithmic law $T\sim\ln{N}$. We have encountered this regime already in the particular case of the prisoner's dilemma specified by  $\Tt-R=P-S\equiv c >0$. These two Darwinian regimes are separated by two regimes of neutral evolution, shown in grey in Fig.~\ref{fig:C}. First, for small $N$ and small differences in the payoffs (i.e., around the point where the four types of games coincide) a $T\sim N$ behavior emerges. Second, at the lines where the snowdrift game turns into the prisoner's dilemma resp. by-product mutualism, the mean extinction time increases as a square-root in the population size, $T\sim\sqrt{N}$. 

Similar to the prisoner's dilemma, we now aim at identifying the edge of neutral evolution, i.e., the crossover from the Darwinian regimes to the regimes of neutral evolution. We have calculated the boundaries of both neutral regimes, $T\sim N$ and $T\sim \sqrt N$ analytically, see Methods section~\ref{methods_edges}. They are described by straight lines for the first one and by parabola-shaped lines for the second one, see Fig.~\ref{fig:C}.

Both edges of neutral evolution scale proportional to the system size $N$. Therefore, while increasing the system size changes the payoff parameters where the crossovers appear,  the shape and relations of the different regimes are not altered. Concerning the dependence of the edges of neutral evolution on the characteristic strength of selection $s$, meaning the average contribution of the fitness-dependent payoff to the overall fitness, different scaling laws arise. 
For the crossover from the neutral regime $T\sim N$ to the other regimes,  $T_e$ and $N_e$ scale as $1/s$. In contrast, a scaling law  $N_e\sim 1/s^2$ for crossovers between the neutral regime with $T\sim \sqrt{N}$ and the Darwinian regimes emerges. This different scaling behavior arises, for example, for $\Tt-\Rr=1$ and varying $s=\Ss-\Pp$ as shown in Fig.~\ref{fig:D} B.

\section{Discussion}

Cooperation is often threatened by exploitation and therefore, although beneficial, vulnerable to extinction. In evolutionary dynamics, this mechanism comes in through selection by individuals' fitness, the driving force of Darwinian evolution.  However, evolution also possesses stochastic aspects. Employing a standard formulation of social dilemmas, we have shown that fluctuations can support cooperation in two distinct ways. First, they can lead cooperators to fully take over the population. Second, neutral evolution considerably increases the time at which cooperators and defectors coexist, i.e., at which a certain level of cooperation is maintained.
To emphasize the importance of the second point, we note that in real ecological systems the rules of the dynamics themselves change due to external~\cite{Raup} or internal~\cite{Levin} influences, setting
an upper limit to the time-scales at which evolution with constant payoffs, as we study here, applies. In particular, these times can be shorter than the times that would be needed for extinction of either cooperators or defectors, such that it may be less important to look at which of both would ultimately remain, but what the time-scales for extinction are.  

Quantitatively, we have shown the emergence of different Darwinian and neutral regimes. In the Darwinian regime of the prisoner's dilemma, cooperators are guaranteed to become extinct; the same is true for the second neutral regime, where $T\sim\sqrt{N}$. However, in the other neutral regime, with $T\sim N$, a random process determines whether cooperators or defectors prevail. Cooperators may therefore take over due to essentially neutral evolution. Moreover, even if cooperators eventually disappear, they remain for a considerably longer time in the neutral regimes than in the Darwinian regime. Indeed,  in the regimes of neutral evolution, coexistence of cooperators and defectors is maintained for a mean time $T$ obeying $T\sim N$ resp. $T\sim\sqrt{N}$. For medium and large population sizes, this time exceeds by far the time $T\sim \ln N$ at which cooperation disappears in the Darwinian regimes of the prisoner's dilemma or of the coordination game (if defectors happen to dominate in the latter case). Neutral evolution can therefore maintain cooperation on a much longer time-scale than Darwinian evolution.  This effect is relevant as the neutral regimes considerably extend into the prisoner's dilemma as well as the cooperation game region. There, a form of neutrally maintained cooperation evolves.

Our results have been obtained by applying a general concept based on extinction times that allows to classify evolutionary dynamics into regimes of Darwinian and neutral character, separated by an emerging edge of neutral evolution. Apart from the social dilemmas under consideration here, we believe that our quantitative analytical approach  can be versatilely applied to disentangle the effects of selection and fluctuations in various ecological situations where different species coexist~\cite{Szabo,Reichenbach2,reichenbach-2008-254,reichenbach-2008-101,May}. Encouraged by our findings, we expect such studies to reveal further unexpected effects of fluctuations on ecology and evolution.


Financial support of the German Excellence Initiative via the program 'Nanosystems Initiative Munich' and the German Research Foundation via the SFB TR12 'Symmetries and Universalities in Mesoscopic Systems'  is gratefully acknowledged. T. R. acknowledges funding by the Elite-Netzwerk Bayern.

\section*{References}


\begin{thebibliography}{10}

\bibitem{Axelrod}
R.~Axelrod.
\newblock {\em The {E}volution of {C}ooperation}.
\newblock Basic Books, New York, 1984.

\bibitem{NowakEGT}
M.~A. Nowak.
\newblock {\em Evolutionary Dynamics: Exploring the Equations of Life}.
\newblock Belknap Press, Cambridge, 2006.

\bibitem{Ratnieks}
F.~L.~W. Ratnieks, K.~R. Foster, and T.~Wenseleers.
\newblock Conflict resolution in insect societies.
\newblock {\em Ann. Rev. Entomol.}, 51:581--608, 2006.

\bibitem{Diggle}
S.~P. Diggle, A.~S. Griffin, G.~S. Campbell, and S.~A. West.
\newblock Cooperation and conflict in quorum-sensing bacterial populations.
\newblock {\em Nature}, 450:411--414, 2007.

\bibitem{Fehr}
E.~Fehr and U.~Fischbacher.
\newblock The nature of human altruism.
\newblock {\em Nature}, 425:785--791, 2003.

\bibitem{axelrod-1981-211}
R.~Axelrod and WD. Hamilton.
\newblock {The evolution of cooperation}.
\newblock {\em Science}, 211:1390--1396, 1981.

\bibitem{milinski-1987-325}
M.~Milinski.
\newblock Tit for tat in sticklebacks and the evolution of cooperation.
\newblock {\em Nature}, 325:433 -- 435, 1987.

\bibitem{rockenbach-2006-444}
B.~Rockenbach and M.~Milinski.
\newblock The efficient interaction of indirect reciprocity and costly
  punishment.
\newblock {\em Nature}, 444:718--723, 2006.

\bibitem{TraulsenGroup}
A.~Traulsen and M.~A. Nowak.
\newblock {Evolution of cooperation by multilevel selection}.
\newblock {\em Proc. Nat. Acad. Sci. USA}, 103(29):10952--10955, 2006.

\bibitem{Szabo2}
G.~Szab\'o, J.~Vukov, and A.~Szolnoki.
\newblock Phase diagrams for an evolutionary prisoner's dilemma game on
  two-dimensional lattices.
\newblock {\em Phys. Rev. E}, 72(4):047107, 2005.

\bibitem{Chuang01092009}
J.~S. Chuang, O.~Rivoire, and S.~Leibler.
\newblock Simpson's paradox in a synthetic microbial system.
\newblock {\em Science}, 323:272--275, 2009.

\bibitem{Fisher}
R.~A. Fisher.
\newblock {\em The Genetical Theory of Natural Selection}.
\newblock Oxford University Press, Oxford, 1930.

\bibitem{Wright}
S.~Wright.
\newblock Evolution in mendelian populations.
\newblock {\em Genetics}, 16:97--159, 1931.

\bibitem{Maynard}
J.~{Maynard Smith}.
\newblock {\em Evolution and the Theory of Games}.
\newblock Cambridge University Press, Cambridge, 1982.

\bibitem{NowakCooperation}
M.~A. Nowak.
\newblock Five rules for the evolution of cooperation.
\newblock {\em Science}, 314:1560, 2006.

\bibitem{Hamilton64}
W.~D. Hamilton.
\newblock The genetical evolution of social behaviour. {I+II}.
\newblock {\em J.Theor. Biol.}, 7:1--52, 1964.

\bibitem{Hamilton}
W.D. Hamilton.
\newblock {\em Narrow Roads of Gene Land: Evolution of Social Behaviour}.
\newblock Oxford University Press, Oxford, 1996.

\bibitem{Trivers}
R.L. Trivers.
\newblock The evolution of reciprocal altruism.
\newblock {\em Quart. Rev. Biol.}, 46:35, 1971.

\bibitem{NowakSigmundTFT}
M.~A. Nowak and K.~Sigmund.
\newblock Tit for tat in heterogeneous populations.
\newblock {\em Nature}, 335:250--253, 1992.

\bibitem{hauert-2007-316}
C.~Hauert, A.~Traulsen, H.~Brandt, M.~A. Nowak, and K.~Sigmund.
\newblock Via freedom to coercion: The emergence of costly punishment.
\newblock {\em Science}, 316:1905--1907, 2007.

\bibitem{Volkov}
I.~Volkov, J.~R. Banavar, S.~P. Hubbell, and A.~Maritan.
\newblock Patterns of relative species abundance in rainforests and coral
  reefs.
\newblock {\em Nature}, 450:45--49, 2007.

\bibitem{Bell}
G.~Bell.
\newblock Neutral macroecology.
\newblock {\em Science}, 293:2413--2418, 2001.

\bibitem{Hubbell}
S.~P. Hubbell.
\newblock {\em The Unified Neutral Theory of Biodiversity and Biogeography}.
\newblock Princeton University Press, Princeton, 2001.

\bibitem{Muneepeerakul}
R.~Muneepeerakul, E.~Bertuzzo, H.~J. Lynch, W.~F. Fagan, A.~Rinaldo, and
  I.~Rodriguez-Iturbe.
\newblock Neutral metacommunity models predict fish diversity patterns in
  {M}ississippi-{M}issouri basin.
\newblock {\em Nature}, 453:220--223, 2008.

\bibitem{Muneepeerakul2}
R.~Muneepeerakul, J.~S. Weitz, S.~A. Levin, A.~Rinaldo, and
  I.~Rodriguez-Iturbe.
\newblock A neutral metapopulation model of biodiversity in river networks.
\newblock {\em J. Theor. Biol.}, 245:351--363, 2007.

\bibitem{nowak-2004-428}
M.~A. Nowak, A.~Sasaki, C.~Taylor, and D.~Fudenberg.
\newblock Emergence of cooperation and evolutionary stability in finite
  populations.
\newblock {\em Nature}, 428:646 -- 650, 2004.

\bibitem{Traulsen}
Arne Traulsen, Jens~Christian Claussen, and Christoph Hauert.
\newblock Coevolutionary dynamics: from finite to infinite populations.
\newblock {\em Phys. Rev. Lett.}, 95:238701, 2005.

\bibitem{Antal}
T.~Antal and I.~Scheuring.
\newblock Fixation of strategies for an evolutionary game in finite
  populations.
\newblock {\em Bull. Math. Biol.}, 68:1923, 2006.

\bibitem{TaylorEx}
C.~Taylor, Y.~Iwasa, and M.A. Nowak.
\newblock A symmetry of fixation times in evolutionary dynamics.
\newblock {\em J. Theor. Biol.}, 243:245--251, 2006.

\bibitem{Moran}
P.~A. Moran.
\newblock {\em The {S}tatistical {P}rocesses of {E}volutionary {T}heory}.
\newblock Clarendon Press Oxford, Oxford, 1964.

\bibitem{Dawes}
R.~M. Dawes.
\newblock Social dilemmas.
\newblock {\em Ann. Rev. Psychol.}, 31:169--193, 1980.

\bibitem{Reichenbach}
T.~Reichenbach, M.~Mobilia, and E.~Frey.
\newblock Mobility promotes and jeopardizes biodiversity in rock-paper-scissors
  games.
\newblock {\em Nature}, 448:1046--1049, 2007.

\bibitem{Cremer}
J.~Cremer, T.~Reichenbach, and E.~Frey.
\newblock Anomalous finite-size effects in the {B}attle of the {S}exes.
\newblock {\em Eur. Phys. J. B}, 63:373--380, 2008.

\bibitem{reichenbach-2006-74}
T.~Reichenbach, M.~Mobilia, and E.~Frey.
\newblock Coexistence versus extinction in the stochastic cyclic
  {L}otka-{V}olterra model.
\newblock {\em Phys. Rev. E}, 74:051907, 2006.

\bibitem{berr:048102}
M.~Berr, T.~Reichenbach, M.~Schottenloher, and E.~Frey.
\newblock Zero-one survival behavior of cyclically competing species.
\newblock {\em Phys. Rev. Lett.}, 102(4):048102, 2009.

\bibitem{traulsen_extime}
P.~M. Altrock and A.~Traulsen.
\newblock Fixation times in evolutionary games under weak selection.
\newblock {\em New J. Phys}, 11(1):013012, 2009.

\bibitem{ewens}
Warren~J. Ewens.
\newblock {\em Mathematical Population Genetics}.
\newblock Springer, New York, 2nd. edition, 2004.

\bibitem{Kimura}
M.~Kimura.
\newblock {\em The Neutral Theory of Molecular Evolution}.
\newblock Cambridge University Press, Cambridge, 1983.

\bibitem{Gardiner}
C.~W. Gardiner.
\newblock {\em Handbook of Stochastic Methods}.
\newblock Springer, Berlin, 2007.

\bibitem{Risken}
H.~Risken.
\newblock {\em The {F}okker-{P}lanck {E}quation - {M}ethods of {S}olution and
  {A}pplications}.
\newblock Springer-Verlag, Heidelberg, 1984.

\bibitem{Kampen}
N.G.~Van Kampen.
\newblock {\em Stochastic Processes in Physics and Chetry (North-Holland
  Personal Library)}.
\newblock North Holland, 2nd edition, 2001.

\bibitem{KimuraFixation}
M.~Kimura.
\newblock Genetic variability maintained in a finite population due to
  mutational production of neutral and nearly neutral isoalleles.
\newblock {\em Genet. Res., Camb.}, 11:247--69, 1969.

\bibitem{Ohta}
T.~Ohta and J.~H. Gillespie.
\newblock Development of neutral and nearly neutral theories.
\newblock {\em Theor. Pop. Biol.}, 49:128--142, 1996.

\bibitem{Raup}
D.~M. Raup.
\newblock The role of extinction in evolution.
\newblock {\em Proc. Nat. Acad. Sci. USA}, 91:6758--6763, 1994.

\bibitem{Levin}
L.~Worden and S.~A. Levin.
\newblock Evolutionary escape from the prisoner's dilemma.
\newblock {\em J. Theor. Biol.}, 245:411--422, 2007.

\bibitem{Szabo}
G.~Szab\'o and G.~F\'ath.
\newblock Evolutionary games on graphs.
\newblock {\em Phys. Rep.}, 446(4-6):97--216, 2007.

\bibitem{Reichenbach2}
T.~Reichenbach, M.~Mobilia, and E.~Frey.
\newblock Noise and correlations in a spatial population model with cyclic
  competition.
\newblock {\em Phys. Rev. Lett.}, 99:238105, 2007.

\bibitem{reichenbach-2008-254}
T.~Reichenbach, M.~Mobilia, and E.~Frey.
\newblock Self-organization of mobile populations in cyclic competition.
\newblock {\em J. Theor. Biol.}, 254:368--383, 2008.

\bibitem{reichenbach-2008-101}
T.~Reichenbach and E.~Frey.
\newblock Instability of spatial patterns and its ambiguous impact on species
  diversity.
\newblock {\em Phys. Rev. Lett.}, 101:058102, 2008.

\bibitem{May}
R.~M. May.
\newblock {\em Stability and Complexity in Model Ecosystems}.
\newblock Princeton Univ. Press, Princeton, 2nd edition, 1974.

\end{thebibliography}

\end{document}